# Bulk-hole correspondence and inner robust boundary modes in singular flatband lattices

Limin Song[1†], Shenyi Gao[1†], Shiqi Xia[1], Yongsheng Liang[1], Liqin Tang[1,2], Daohong Song[1,2*],
Daniel Leykam[3], and Zhigang Chen[1,2*]

[1]*The MOE Key Laboratory of Weak-Light Nonlinear Photonics, TEDA Applied Physics Institute and School of Physics, Nankai University, Tianjin 300457, China*
[2]*Collaborative Innovation Center of Extreme Optics, Shanxi University, Taiyuan, Shanxi 030006, China*
[3]*Science, Mathematics and Technology Cluster, Singapore University of Technology and Design, 8 Somapah Road, 487372 Singapore*
[†]*These authors contributed equally to this work*
[*]*songdaohong@nankai.edu.cn, zgchen@nankai.edu.cn*

**Abstract:**

Topological entities based on bulk-boundary correspondence are ubiquitous, from conventional to higher-order topological insulators, where the protected states are typically localized at the outer boundaries (edges or corners). A less explored scenario involves protected states that are localized at the inner boundaries, sharing the same energy as the bulk states. Here, we propose and demonstrate what we refer to as the *bulk-hole correspondence* - a relation between the inner robust boundary modes (RBMs) and the existence of multiple "holes" in singular flatband lattices, mediated by the immovable discontinuity of the bulk Bloch wavefunctions. We find that the number of independent flatband states always equals the sum of the number of independent compact localized states and the number of nontrivial inner RBMs, as captured by the Betti number that also counts the hole number from topological data analysis. This correspondence is universal for singular flatband lattices, regardless of the lattice shape and the hole shape. Using laser-written Kagome lattices as a platform, we experimentally observe such inner RBMs, demonstrating their real-space topological nature and robustness. Our results may extend to other singular flatband systems beyond photonics, including non-Euclidean lattices, providing a new approach for understanding nontrivial flatband states and topology in hole-bearing lattice systems.

**Keywords:** Singular flat band, robust boundary mode, Betti number, real-space topology, bulk-hole correspondence, Kagome lattice

Topological bulk-boundary correspondence has received a great deal of attention, partly due to its importance for understanding different topological insulators (TIs) including the Chern insulators and higher-order TIs [1-5]. In those cases, a topological state is typically localized at the system's boundaries, protected by a bulk topological invariant. In a different scenario, a topological state may reside inside the bulk structure. To this end, a simple idea is to create an ''inner boundary'' (such as through a ''hole'', or a topological defect) inside a lattice that can support such a state. A variety of TIs with interesting geometries have been proposed, for example in the study of nonlinearity-driven photonic Floquet TIs [6], fractal TIs [7, 8], as well as topological crystalline insulators with disclinations [9-14]. Those geometries and implementations often involve complex design and fabrication. On the other hand, it has been demonstrated that singular flatband (SFB) systems [15, 16] can also support topological entities [17-22], although the topological features in those systems arise from real space rather than momentum space.

Singular flat bands are identified by the immovable discontinuity of the Bloch wavefunctions [15, 16], characterized by a flatband touching with adjacent dispersive band(s). As a typical example, the Kagome lattices [15, 16, 23-29] (and related SFB lattices from line graphs [30]) feature a representative geometry that possesses one SFB in the band structure [Figs. 1(a) and 1(b)]. SFB systems exhibit numerous intriguing phenomena, including noncomplete flatband spanning sets constructed by compact localized states (CLSs) [31-33], noncontractible loop states (NLSs) protected by real-space topology [18, 19, 34], and those related to nonzero quantum distance and anomalous Landau levels [35, 36]. The NLSs - a type of real-space topological entity, can be considered as line states along $a_1$ and $a_2$ directions [Fig. 1(a)] in an infinite 2D lattice, or loop states mapped to the poloidal and toroidal directions in a torus geometry [Fig. 1(e)] [15, 34]. The first direct observation of such NLSs was realized in a Kagome lattice with Corbino-shaped geometry [18]. Recent studies have included non-Euclidean lattices [37, 38], lattices with co-existing singular and nonsingular flat bands [19], and those with partially flat bands [21]. The increased attention on NLSs mainly stems from two aspects: One is their origin from the SFB touching point in momentum space; the other is their topological protection from real space [15, 34, 37]. For the former aspect, it is often useful to probe the NLSs and unconventional line states [17, 39] so to characterize the SFB touching. However, such an approach requires either periodic boundary conditions or specific tailored boundaries not readily applicable for more generalized cases. For the latter aspect, currently there is no real-space topological invariant to describe the underlying physics of the SFB systems. Therefore, an important yet challenging task is to find an intuitive and efficient way to identify the SFB characteristics and study underlying real-space topology. The so-called robust boundary modes (RBMs) [15] have emerged as a promising candidate, although such boundary-localized modes [Fig. 1(c)] are considered only as "indirect" manifestations of the NLSs [18, 19]. Conventionally, the RBMs refer to *outer* boundary states resulting from vanishing bulk Bloch wavefunction at the SFB touching point, and they are also considered as the real-space topological entity [15, 16].

In this work, we report the realization of *inner* RBMs that illustrate the interplay of real-space topology and nontrivial flatband lattices, characterized by the first Betti number ($\mathcal{B}_1$) which counts the number of lattice holes hosting the inner RBMs. By mapping each lattice site to a data point, we construct a data set for an SFB lattice and obtain corresponding $\mathcal{B}_1$ by using the method of topological data analysis. Then, we establish a *bulk-hole correspondence* (BHC), which links the vanishing bulk Bloch wavefunction to the inner RBMs counted by $\mathcal{B}_1$. Experimentally, by employing a continuous-wave laser-writing technique, we create a photonic Kagome lattice with hole structures, and observe the inner RBMs with the state number equal to $\mathcal{B}_1$, thus demonstrating the BHC. We show that, inherited from the NLSs, the inner RBMs have a real-space topological origin. In fact, an inner RBM cannot be disconnected, but it can be transformed into an outer counterpart by superposition of the bulk CLSs featured by the faltband lattices.

We first give a brief introduction to the concepts of the hole number and the Betti number used in this work. A hole is functionally equivalent to a dislocation or disclination in a lattice [40-43], as it can be considered as one kind of real-space topological defect. Two-dimensional lattices (or corresponding abstract geometrical shapes) can be topologically classified by their Betti numbers, which are widely used in topological data analysis for identification of topological features of abstract shapes [44-46]. The $n$-th Betti number $\mathcal{B}_n$ refers to the number of $n$-dimensional ($n$D) holes on a topological surface. The first few Betti numbers are defined for 0D, 1D, and 2D simplicial complexes as follows [46]: $\mathcal{B}_0$ is the number of connected components; $\mathcal{B}_1$ is the number of 1D or "circular" holes; $\mathcal{B}_2$ is the number of 2D "voids" or "cavities" [45]. For instance, a plane without a hole has $\mathcal{B}_1 = 0$ [Fig. 1(f)], while a plane with one hole has $\mathcal{B}_1 = 1$ [Fig. 1(g)] - see calculation details about the Betti numbers in Supplementary Material (SM).

Next, we show both the inner and outer RBMs can be related to the SFB touching. We consider a Kagome lattice consisting of weakly coupled waveguide arrays that can be described by the discrete tight-binding model [47]. The flatband Bloch wavefunction of the Kagome lattice can be readily obtained as

$$|\psi(\boldsymbol{k})\rangle = \frac{1}{\alpha_{\boldsymbol{k}}} \left( e^{i\boldsymbol{k}\cdot\boldsymbol{a}_1} - 1, 1 - e^{i\boldsymbol{k}\cdot(\boldsymbol{a}_1-\boldsymbol{a}_2)}, e^{i\boldsymbol{k}\cdot(\boldsymbol{a}_1-\boldsymbol{a}_2)} - e^{i\boldsymbol{k}\cdot\boldsymbol{a}_1} \right)^{\mathrm{T}}, \qquad (1)$$

where $\boldsymbol{k} = (k_x, k_y)$ is the momentum and $\alpha_{\boldsymbol{k}} = \sqrt{2}\sqrt{3 - \cos(\boldsymbol{k}\cdot\boldsymbol{a}_1) - \cos(\boldsymbol{k}\cdot\boldsymbol{a}_2) - \cos[\boldsymbol{k}(\boldsymbol{a}_1-\boldsymbol{a}_2)]}$ is a normalization factor. By applying an inverse Fourier transformation to Eq. (1), one can obtain a typical

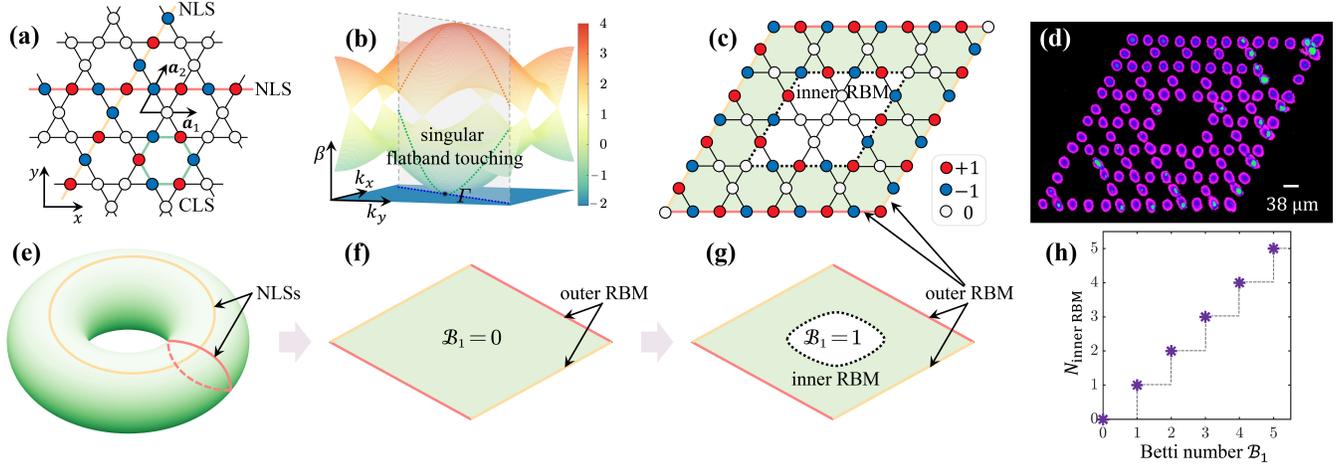

Fig. 1: **Schematic illustration and band structure of a Kagome lattice and its associated RBM and NLS.** (a) Illustration of a Kagome lattice, where $\boldsymbol{a}_1 = a(1,0)$ and $\boldsymbol{a}_2 = a(1/2, \sqrt{3}/2)$ are the two lattice vectors, and $a$ is the lattice constant. One compact localized state (CLS) is marked by a green hexagon, and two noncontractible loop states (NLSs) marked in red and yellow are located along $\boldsymbol{a}_1$ and $\boldsymbol{a}_2$ directions. (b) Calculated band structure in which a flat band touches one of the dispersive bands at the high-symmetry point Γ(0,0), resulting in a singular flat band. The gray region is a 2D cross-sectional view of the band structure at the flatband touching point. (c) The outer and inner robust boundary modes (RBMs) are connected by colored solid lines and black dashed lines, respectively. (d) A representative experimental lattice with a Betti number $\mathcal{B}_1 = 3$. (e) The two NLSs in (a) are mapped onto a torus, representing a Kagome lattice with periodic boundary condition. They span the yellow and pink closed loops winding around the torus, thereby manifesting nontrivial real-space topology. (f) The NLSs in (e) become an outer RBM when the torus is "cut" into open boundary. (g) The open-boundary system obtained from (f) by adding a hole, where an inner RBM (dashed line) is created. Different Betti numbers $\mathcal{B}_1 = 1$ and $\mathcal{B}_1 = 0$ distinguish the two cases (with and without a hole). (h) The number of the inner RBMs increases as a function of $\mathcal{B}_1$. In (a) and (c), white empty lattice sites represent zero amplitude, while filled blue and red ones represent nonzero equal amplitude but with opposite phase.

CLS with equal amplitudes and alternating phase between nearest-neighbor sites in a hexagonal plaquette displayed in Fig. 1(a). Other CLSs residing on different plaquettes are translated duplications of this CLS. The immovable discontinuity (SFB touching point) at $\boldsymbol{k} = (0,0)$ makes all the components of Eq. (1) vanish [15, 16], leading to a vanishing sum of the CLSs under periodic boundary conditions. However, their linear combination gives rise to an outer RBM [the chromatic solid lines in Fig. 1(c, f, g)] encircling a finite-sized lattice under the open boundary condition. As illustrated in Fig. 1(c), one can introduce a hole to the lattice and then obtain an inner RBM. The outer and inner RBMs share the same alternating phase structure, beneficial for destructive interference into the bulk. Interestingly, if a specific relative phase is assumed, the inner and outer RBMs are interconvertible through linear combinations of the bulk CLSs; however, their relative phase can be freely tunable for a big lattice (see SM) [15, 20]. An abstract geometrical representation for this procedure can be modeled in Fig. 1(g), where the resulting inner RBMs are located at the boundary of the holes characterized by $\mathcal{B}_1 = 1$.

Theoretically, we design several groups of lattice geometries with different hole numbers to illustrate the BHC. In Fig. 2(a), a Kagome lattice with a rhombic shape without holes, i.e., $\mathcal{B}_1 = 0$, is considered. Calculated eigenvalue spectra show that the number of independent flatband states ($N_{\text{FBS}}$) is the same as the number of the basic CLSs ($N_{\text{CLS}}$), or equivalently, the number of the plaquettes. This simply indicates that there are no extra flatband states linearly independent of the CLSs. However, when one hole is introduced to the lattice by removing some interconnected sites [Fig. 2(b)], i.e., with $\mathcal{B}_1 = 1$, a fundamental change occurs in the underlying flatband states: $N_{\text{FBS}}$, albeit smaller compared with the case $\mathcal{B}_1 = 0$, is no longer equal to $N_{\text{CLS}}$. Interestingly, the former is exactly one more than the latter. This is independent of specific shape and/or size of the lattice, as well as the number of removed sites at the inner boundary (see SM for more examples including triangle- and hexagon-shaped Kagome lattices).

An explicit feature emerges as we continue to add more holes to the lattices. In Figs. 2(b) and 2(c), two and three holes are introduced, labeled by $\mathcal{B}_1 = 2,3$, respectively. Results show again that $N_{\text{FBS}}$ is greater than $N_{\text{CLS}}$ by a number equal to $\mathcal{B}_1$. Moreover, $\mathcal{B}_1$ remains independent of specific hole shapes. Through extensive calculations, we establish a universal relation, i.e., the BHC, which can be formulated as

$$N_{\text{FBS}} = N_{\text{CLS}} + \mathcal{B}_1, \ (\text{where } \mathcal{B}_1 = N_{\text{inner RBM}}) \tag{2}$$

where $N_{\text{inner RBM}}$ describes the number of the inner RBMs (see more discussions on the BHC in SM). We find that $N_{\text{inner RBM}}$ is solely determined by $\mathcal{B}_1$ and shows a stepwise increase with the Betti number [Fig. 1(h)]. In Figs. 2(b-c) we display these inner RBMs, which reside on the boundaries of holes, and assign alternating amplitudes to nearest-neighbor sites. It is well known that the basic CLSs localized at different hexagonal plaquettes are linearly independent to each other [15, 34, 37]. Similarly, inner RBMs can be considered as the boundary localization from "combined plaquettes". Thus, the inner RBMs localized at different holes are linearly independent of each other, as counted by $\mathcal{B}_1$.

We now illustrate the real-space topological equivalence between perturbed sublattices (realized by applying on-site disorders $\delta_i$, and $i = 1,2,\cdots,N$ denote the indexes of the perturbed sublattices or defect sites) and actual holes introduced in finite-sized SFB lattices. Specifically, the perturbed sublattices form an effective "hole" by connecting the plaquettes directly associated with them. For example, in Fig. 2(d), the perturbed sublattices with equal and unequal $\delta_i$ surrounded by each dashed line form three holes characterized by $\mathcal{B}_1 = 3$.

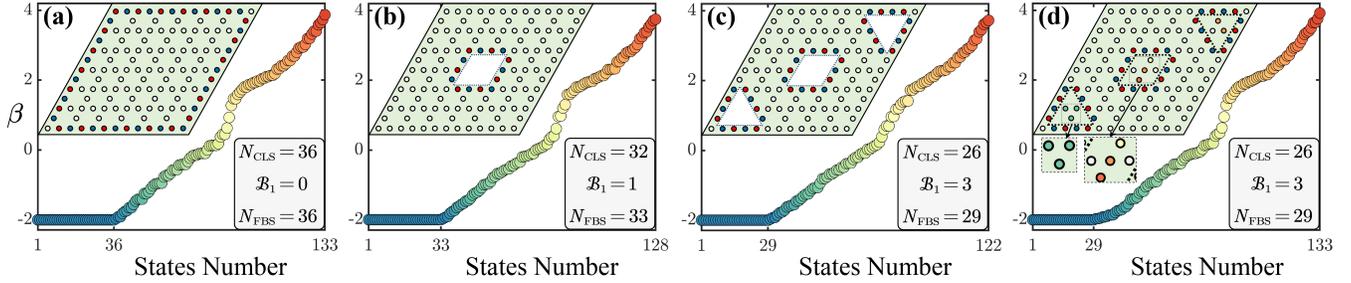

Fig. 2: **Typical examples of shaped Kagome lattices used to demonstrate bulk-hole correspondence (BHC).** All subplots (a)-(d) have the same layout: Top-left insets display rhombic lattice geometries with different numbers of "holes" counted by the Betti number $\mathcal{B}_1$; Main plots along off-diagonal direction show the eigenvalue spectra of corresponding lattices, for which the colormap is the same as that in Fig. 1(b); Bottom-right insets list the number of independent compact localized states $N_{\text{CLS}}$, the $\mathcal{B}_1$, and the number of total flatband states $N_{\text{FBS}}$ (labeled on the horizontal axis). One can easily find that $N_{\text{FBS}} = N_{\text{CLS}} + \mathcal{B}_1$ always holds. We show the amplitude and phase profiles of the outer (inner) RBMs for the $\mathcal{B}_1 = 0$ ($\mathcal{B}_1 \neq 0$) cases with colored dots in each lattice structure. In (d), the empty sites are replaced by differently colored sites within the dashed regions, representing different on-site potentials as perturbations of different strengths $\delta_i \in [-2,4]$, where $i = 1,2,\cdots,N$ counts the number of sites from left to right along a given line. The zoom-in insets in (d) are for better illustration of different cases with equal and unequal $\delta_i$, respectively.

Next, we present the experimental observation of these inner RBMs using a photonic platform. The BHC can be examined by probing the inner RBMs and identifying their number equal to $\mathcal{B}_1$, since it has been shown that the RBMs stem from the singularity of the Bloch wavefunction [15, 16, 20]. Without loss of generality, we take a rhombic-shaped Kagome lattice with 3 holes shown in Fig. 2(c) as a typical example. Such a lattice with $\mathcal{B}_1 = 3$ is established in a nonlinear (SBN:61) crystal via laser-writing [17] [Fig. 1(d)]. In this experiment, a Gaussian-shaped writing beam (with input power 2.5 μW, wavelength 488 nm, and full width at half maximum 8 μm) is employed to induce waveguides one-by-one in the crystal, establishing a nearly uniform lattice of about 38 μm spacing. Our tight-binding model works well with such lattice parameters in experiment. To observe the inner RBMs, a specially designed probe beam is launched into the lattice, assisted with a spatial light modulator so the probe beam has its intensity [Fig.

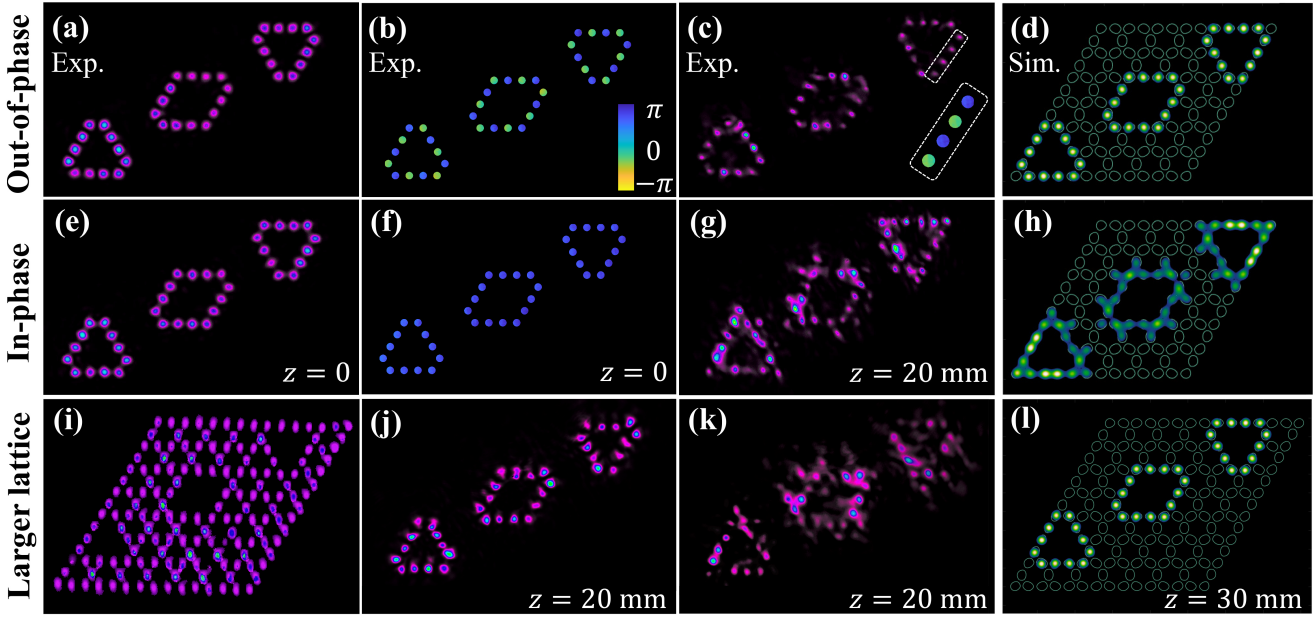

Fig. 3: **Experimental demonstration of inner RBMs and the BHC.** The first set of experimental results (a-h) are obtained from a 3-hole laser-written lattice shown in Fig. 1(d), where the lattice has 122 sites in total. (a) Input intensity patterns of a probe beam matching the mode distribution of three inner RBMs shown in Fig. 2(c), with the lattice Betti number $\mathcal{B}_1 = 3$. (b) Phase distribution of the probe beam, showing out-of-phase relation between adjacent sites. (c) Output patterns of (a) after 20-mm-long propagation through the written lattice. Dashed rectangles illustrate the measured output phase relation of the selected sites, confirming opposite phase between adjacent sites. Note that the three inner RBMs can be tested one by one or all together at once. (d) Simulated results under the same experimental condition. (e)-(h) Results have the same layout as (a)-(d), but are obtained from in-phase excitation (i.e., all sites have equal phase at input) for a direct comparison, showing initial excitation cannot be localized at the inner boundaries. (i)-(k) Another set of experimental results involving a larger number of lattice sites (165) are shown in the bottom panel, together with long distance simulation (l) to show the robustness of the localized inner RBMs.

3(a)] and phase [Fig. 3(b)] reconfigured to match the inner RBMs in Fig. 2(c). The phase relation of the excited sites is extracted from the interferogram of the probe beam by applying the Fourier transformation [48]. After 20-mm-long propagation through the lattice, the output intensity pattern [Fig. 3(c)] of the probe beam remains localized at the initially excited waveguides outlining the three holes. In comparison, as the in-phase probe beam [Figs. 3(e) and 3(f)] is launched into the same lattice sites, the output pattern is strongly distorted and exhibits evident coupling to other sites [Fig. 3(g)] due to discrete diffraction. An obvious phenomenon is the corner-site intensity distribution at the bottom-left and top-right corners of the Kagome lattice [Fig. 3(g)], but it should be noted that there is no corner state in our structure. These experimental results of the inner RBMs are corroborated by numerical simulations [Figs. 3(d) and 3(h)] using parameters from the experiment.

The BHC is independent of the size and shape of the lattice. To illustrate this in experiment, we construct another lattice with the number of bulk lattice sites increased to 165, but with the same Betti number $\mathcal{B}_1 = 3$ [Fig. 3(i)]. From direct comparison of the out-of-phase [Fig. 3(j)] and in-phase [Fig. 3(k)] outputs (excitations are the same as Figs. 3(a,b) and 3(e,f), respectively), together with long distance simulation [Fig. 3(l)], we show the strong and compact localization of the inner RBMs. These results further prove that the BHC is independent of the lattice size.

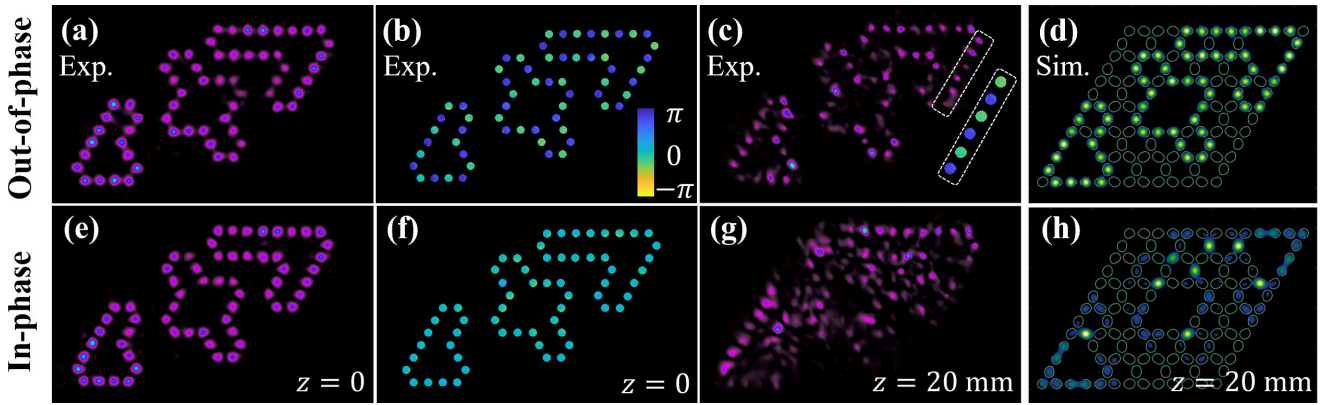

Fig. 4: **Experimental demonstration of inner RBM's robustness under perturbation from intruding CLSs.** (a)-(h) Results are obtained under same lattice conditions as Figs. 3(a)-3(h), except that the probe beam has been reconfigured by adding additional CLSs. Specifically, for the three inner RBMs shown in Fig. 3, one, two, and three CLSs are intentionally interfered in the initial polygon-shaped intensity patterns in the bottom-left, middle, and top-right inner RBMs, respectively. Although the symmetry of the original inner RBMs is no longer present, these arbitrarily-shaped RBMs remain intact during propagation.

The RBMs are proposed to exhibit distinctive features that are robust against defects and perturbations. Specifically, their boundary loops cannot be cut or disconnected by adding/subtracting a finite number of CLSs [15], even though the loops and the shape of the RBMs are strongly deformed. From this perspective, we perform another set of experiments to illustrate the robust property of the inner RBMs. As shown in Fig. 4, the intensity and phase profiles of the probe beam are intentionally modulated at input [Figs. 4(a) and 4(b)], such that the probe does not have a regular polygon shape as those inner modes depicted in Fig. 2(c). We observe that its output pattern remains confined to the initially injected lattice sites without spreading into other unexcited sites [Fig. 4(c)], independent of the added CLSs preserving the symmetry of the original inner RBMs or not. In contrast, for in-phase excitation [Figs. 4(e)-4(f)], light spreads to other bulk sites out of the initially excited region, making the initial pattern strongly distorted [Fig. 4(g)]. Corresponding numerical simulations are presented in Figs. 4(d)-4(h). These results confirm the robustness of the inner RBMs mediated by real-space topology and the BHC.

In conclusion, we have proposed and experimentally demonstrated an intriguing BHC in SFB lattices wherein nontrivial inner RBMs are counted by the Betti number of an abstract lattice shape. Since the formation of inner RBMs requires less strict boundary conditions than does that of conventional topological edge states, light trapped at holes or defects may be observable in a wide range of platforms beyond waveguide systems. Furthermore, the BHC may inspire future studies on real-space topology in non-Euclidean geometries [38] and fractal-like structures [19, 49] which inherently contain lattice holes. Our scheme based on the Betti number can also be further developed to better classify the SFB lattices through topological data analysis, which is useful for identifying nontrivial states from spectral calculations. This work not only provides a promising approach to realize robust light transport, but also paves the way for exploring exotic flatband states and nontrivial topological phenomena beyond photonic systems.

**Acknowledgments**

This research was supported by the National Natural Science Foundation of China (No. 12134006, 12374309, and 12274242); the National Key R&D Program of China (No. 2022YFA1404800); the Natural Science Foundation of Tianjin (No. 21JCYBJC00060); the Natural Science Foundation of Tianjin for Distinguished Young Scientists (No. 21JCJQJC00050); and the 111 Project (No. B23045) in China.